\documentclass[aip,cha,amsmath,amssymb,amsfonts,nofootinbib,preprint,nobibnotes]{revtex4-1}

\usepackage{graphicx}
\usepackage[utf8]{inputenc}
\usepackage[T1]{fontenc}
\usepackage[colorlinks=true,linkcolor=blue,anchorcolor=blue,citecolor=blue,filecolor=blue,urlcolor=blue,breaklinks]{hyperref}
\usepackage{xcolor}

\begin{document}

\title{A quality control analysis of the resting state hypothesis via permutation entropy on EEG recordings}

\author{Alessio Perinelli}
\email[]{alessio.perinelli@unitn.it}
\affiliation{Department of Physics, University of Trento, 38123 Trento, Italy}
\affiliation{INFN-TIFPA, University of Trento, 38123 Trento, Italy}
\author{Leonardo Ricci}
\affiliation{Department of Physics, University of Trento, 38123 Trento, Italy}
\affiliation{CIMeC, Center for Mind/Brain Sciences, University of Trento, 38068 Rovereto, Italy}

\date{\today}

\begin{abstract}
The analysis of electrophysiological recordings of the human brain in resting state is a key experimental technique in neuroscience. Resting state is indeed the default condition to characterize brain dynamics. Its successful implementation relies both on the capacity of subjects to comply with the requirement of staying awake while not performing any cognitive task, and on the capacity of the experimenter to validate that compliance. Here we propose a novel approach, based on permutation entropy, to provide a quality control of the resting state condition by evaluating its stability during a recording. We combine the calculation of permutation entropy with a method for the estimation of its uncertainty out of a single time series, thus enabling a statistically robust assessment of resting state stationarity. The approach is showcased on electroencephalographic data recorded from young and elderly subjects and considering both eyes-closed and eyes-opened resting state conditions. Besides showing the reliability of the approach, the results showed higher instability in elderly subjects that hint at a qualitative difference between the two age groups with regard to the distribution of unstable activity within the brain. The method is therefore a tool that provides insights on the issue of resting state stability of interest for neuroscience experiments. The method can be applied to other kinds of electrophysiological data like, for example, magnetoencephalographic recordings. In addition, provided that suitable hardware and software processing units are used, its implementation, which consists here of \emph{a posteriori} analysis, can be translated into a real time one.
\end{abstract}

\maketitle

\begin{quotation}
Countless studies in neuroscience rely on the analysis of experimentally recorded electrophysiological activity of the human brain during the so-called resting state, i.e.\ when subjects are not performing any cognitive task. However, controlling a subject's compliance with this condition is usually not possible during an experiment, and the recorded data might be contaminated by unwanted and unknown activity. In this work, we propose an approach to mitigate this typically overlooked issue by analyzing electroencephalographic data and detecting possible drifts in brain dynamics, corresponding to an unstable resting state condition that compromises the recording's reliability. To carry out this analysis we rely on permutation entropy, an information-theoretical measure of complexity of a time series. A statistically significant drift in permutation entropy along a recording is linked to a lack of stationarity, thus providing a marker of instability of the resting state. Our method makes up a simple and powerful tool to carry out quality control of electrophysiological data, as well as of other time series for which the detection of non-stationarity is essential.
\end{quotation}

\section{Introduction}
\label{sec_introduction}
The characterization of brain states by relying on the analysis of electrophysiological recordings is a fundamental issue in neuroscience and a basic tool in the medical practice. Among the possible brain states, the so-called resting state plays a major role in the investigation of brain functional organization~\cite{RaichlePNAS2001}. The interest in resting state brain dynamics was first prompted by Biswal et al.~\cite{Biswal1995} in the context of functional magnetic resonance imaging (MRI). Since then, the resting state paradigm has been widely used also in connection with electroencephalographic (EEG) and magnetoencephalographic (MEG) analytical techniques~\cite{VanDiessen2015,PerinelliChiariRicci2018,SciRepNSE2019}. A well-known result concerning the brain organization in resting state is provided by the identification, primarily via functional MRI, of resting state networks~\cite{Fox2007,Yeo2011}. Among these, the ``default mode'' network is possibly the most studied one~\cite{Buckner2008,DePasquale2010}.

The definition of the resting state condition is procedural: tipically, resting state corresponds to subjects being instructed by an operator to ``do nothing'', i.e.\ not to perform any cognitive task, while either keeping their eyes closed or fixating a static marker on a screen. Once the instruction is provided, the subject is entrusted with its implementation. As a matter of fact, the operator does not have more than trust to depend on in order to establish the reliability of the experiment outcome. How to objectively determine the quality of the resting state implementation is an unsolved issue. Two different approaches can be envisaged. A first possibility is to devise rigorous and reproducible protocols to be followed when conducting resting state experiments~\cite{Caeyenberghs2024}. Indeed, as some researchers have pointed out~\cite{Lv2018,OConnor2019}, a lack of a shared consensus on the implementation of a resting state is a possible reason for the ongoing underutilization of resting state measurements in clinical applications. However, regardless of how meticulously it is implemented, a protocol cannot provide any \emph{a posteriori} assessment of the reliability of recorded data. Conversely, the second approach consists of assessing the stability of resting state by means of a measurable quantity. This strategy was followed, for example, by using graph metrics in order to evaluate the reliability of resting state brain networks identified through functional MRI~\cite{Andellini2015}.

The goal of the present paper is to address the problem of stability of the resting state assumption by analyzing segments of EEG time series to detect variations in the dynamical state of the brain. To this purpose, permutation entropy~\cite{BandtPompe} (PE) was used as a marker of time series complexity. Indeed, the analysis of EEG recordings has been increasingly tackled by means of approaches typical of information theory: PE, which has become widespread thanks to its simplicity of implementation, was proposed, for example, as a mean to automatically detect epileptic seizure~\cite{Nicolaou2012} and, more recently, as a biomarker for Alzheimer's disease~\cite{Seker2021}.

The data considered here are EEG recordings of healthy young and elderly subjects and reconstructed at 30 brain locations. Data are extracted from the LEMON public database~\cite{Babayan2019}, in which resting state EEG recordings are available both in eyes-closed (EC) and eyes-opened (EO) conditions. The availability of several interleaved EC-EO segments for each subject provided the necessary segmentation to evaluate resting state stability. Although one might argue that alternating EC-EO states violates the pure resting state assignment, the simplicity of the task is not expected to affect the resting state condition, while avoiding mental drifts like falling asleep or boredom.

The analysis proposed here is carried out by evaluating PE on different segments under the null hypothesis of its constancy during the whole recording in each one of the two conditions. The analysis immediately calls for the necessity of estimating the uncertainty of PE assessments. To this purpose, we relied on a recently developed technique~\cite{Entropy2022} that allows for overcoming the problem of each time series being a singleton.

The scope of application of the method proposed in the present work is not limited to neuroscience. Indeed, the approach discussed here is useful in any field where it is necessary to detect nonstationarity of a system's dynamics out of time series.

The present paper is organized as follows. The available dataset, the related preprocessing, as well as a summary of the method for the evaluation PE and its uncertainty are described in Sec.~\ref{sec:methods}. The evaluation of the stability of the resting state for a single subject and node is the topic of Sec.~\ref{sec:assessing_stability}, whereas Sec.~\ref{sec:stability_analysis} addresses the dependence of stability on age, condition and brain location. In Sec.\ref{sec:complexity_stable_resting_states} we briefly address the complexity of stable resting states. Conclusive remarks on the implications of our results are discussed in Sec.~\ref{sec:discussion}.

\section{Materials and methods}
\label{sec:methods}

\subsection{Dataset and preprocessing}
\label{sec:preprocessing}
EEG data used in the present work belong to the Leipzig Mind-Brain-Body ``LEMON'' database~\cite{Babayan2019}, which is publicly available~\cite{LEMONweb}. Data were recorded in compliance with the Declaration of Helsinki and the related study protocol was approved by the ethical committee at the University of Leipzig (reference number 154/13-ff, see also Ref.~\cite{Babayan2019}). The available raw recordings were acquired in a sound-attenuated EEG booth by means of a 62-channels active ActiCAP EEG device whose electrodes were attached according to the international standard 10-20 system. The corresponding signals are digitized with a sampling rate of ${2.5~\text{kHz}}$ upon bandpass-filtering them between ${0.015~\text{Hz}}$ and ${1~\text{kHz}}$. Further details on the data acquisition protocol are available in Ref.~\cite{Babayan2019}. The dataset considered here corresponds to two sets of subjects: a ``young'' set of 30 healthy subjects in the age range between 20 and 35 years old (19 males, 11 females), and an ``elderly'' set of 30 healthy subjects in the age range between 60 and 80 years old (19 males, 11 females).

Data preprocessing and source reconstruction procedure are described in details in Ref.~\cite{Chaos2021} (Appendix B) and in Ref.~\cite{NeuroImage2022} (Sec.~2). For the sake of clarity, we summarize here the key steps. Raw EEG recordings were first filtered within the frequency band between ${0.1~\text{Hz}}$ and ${40~\text{Hz}}$ by relying of fourth-order high-pass and low-pass filters; second, the power line frequency at ${50~\text{Hz}}$ was removed by means of a notch filter; third, the sampling frequency was reduced to ${250~\text{Hz}}$ by downsampling the data. Artifacts due to cardiac, muscular and eye activity were removed by relying on an independent component analysis. To carry out source reconstruction, head models were built out of the individual MRI scans provided in the LEMON database. Source activity was reconstructed by means of the exact low-resolution electromagnetic tomography algorithm (``eLoreta'') that provided current dipoles with unconstrained orientations on a 10 mm template grid. The extracted sequences correspond to current dipole power reconstructed at 30 brain locations (henceforth referred to as ``nodes'') selected as the centroids of 30 brain regions among the 360 defined in the atlas by Glasser \emph{et al.}~\cite{Glasser2016}. The regions considered are V1, V6A, V4 (occipital); 4, 5L, 6mp (central); Pfm, PF, STV (parietal); STGa, TE1a, TA2 (temporal); 10d, 10pp, p10p (frontal); each region was selected symmetrically in both hemispheres.

Each EEG acquisition run consists of 16 alternated segments in EC (8 segments) and EO (8 segments) conditions. To reduce transient effects, we selected 12 consecutive segments, namely 6 EC and 6 EO, by discarding the first and last two segments of each run. Each raw segment covers between ${\gtrsim 60~\text{s}}$ and ${90~\text{s}}$: with the same purpose of reducing transients, we trimmed each segment to ${60~\text{s}}$, corresponding to 15000 samples, by symmetrically removing the leading and trailing data points.

\subsection{Permutation entropy}
\label{sec:pe_assessment}
To compute permutation entropy, $m$-dimensional trajectories ${\mathbf{y}_n}$ are constructed by selecting $m$ consecutive elements from a scalar time series ${Y = \{y_n\}}$: ${\mathbf{y}_n = (y_n, y_{n+1}, \ldots, y_{n+m-1})}$. A trajectory is then encoded into a permutation ${(s_{n,1},\ldots,s_{n,m})}$, were $s_{n,j}$ are integer numbers each corresponding to the rank of $y_{n+j-1}$ within the trajectory $\mathbf{y}_n$. As the permutation elements $s_{n,j}$ belong to the range $[1,m]$, the number of possible permutations is $m!$. For each possible permutation, the observed rate $\hat{p}_S$ is assessed as the occurrence frequency of that permutation. Permutation entropy is then estimated out of the observed rates $\hat{p}_S$ according to the following expression:
\begin{equation*}
	\hat{H}_m(Y) = -\sum_{S}(\hat{p}_S \ln \hat{p}_S) + \frac{\hat{M} - 1}{2(N-m+1)} \,,
\end{equation*}
where the sum corresponds to the so-called plug-in estimator, while the second term is the Miller-Madow correction~\cite{Miller1955,Harris1975} that compensates the plug-in estimator bias and depends on the time series length $N$ and the number $\hat{M}$ of observed permutations ($\hat{M} \leqslant m!$). (We henceforth assume that $0 \ln 0 = 0$).

The growth rate of PE with the dimension $m$ corresponds, asymptotically, to the Kolmogorov-Sinai (KS) entropy of the underlying source. However, because the number of possible permutations grows as $m!$, the evaluation of PE for $m \gtrsim 10$, and thus of the KS entropy, is impractical. Consequently, PE is instead typically employed at fixed $m$ as a marker of complexity or---as in the present work---to detect nonstationarity by evaluating it on different segments of a time series~\cite{Cao2004}. In this context, the demand of a large $m$ has to be traded off against the fact that an unbiased estmation of PE requires the length of the input time series to be much larger than the number of observed permutations. The analysis discussed here was carried out by considering trajectories having dimension $m = 7$.

\subsection{Permutation entropy uncertainty estimation}
\label{sec:pe_uncertainty}
A key requirement of the analysis discussed in the present work is the estimation of the uncertainty associated to each PE assessment. To this purpose, we apply a recently developed method~\cite{Entropy2022} that relies on the construction of a set of proxy time series out of the original one via surrogate generation~\cite{Schreiber2000}. Specifically, the surrogate generation algorithm used here is the iterative amplitude-adjusted Fourier transform (IAAFT) algorithm~\cite{Schreiber1996,JODIChaos2019}, whereby the amplitude distribution and the (approximate) autocorrelation of the original time series are preserved in the surrogate ones. Upon generating a number $L$ of surrogate time series and computing the corresponding PE values, the uncertainty on the PE of the original time series is provided by the standard deviation of the $L$ surrogate PE assessments multiplied times a suitably trimmed scaling factor $\alpha$. For a detailed description of the method, the reader is referred to Ref.~\cite{Entropy2022} (Sec.~3). Following the guidelines provided therein, the number of surrogate time series to be generated for each evaluation was set here to $L = 100$ and the scaling factor $\alpha$ was set to $2$.

\section{Assessing the stability of the resting state}
\label{sec:assessing_stability}
\begin{figure}[h!]
	\centering
	\includegraphics[scale=1.2]{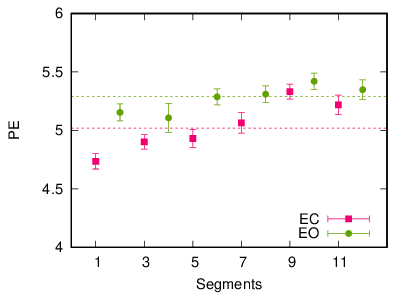}
 	\caption{Example, for a single subject, of the resting state stability analysis. The dashed lines correspond to the constant $\hat{h}$ fitted to the two sets (EC and EO) of PE values. As explained in the main text, the EC condition (red squares) is unstable, whereas the EO condition (green dots) is stable.}
 	\label{fig:fit_example}
\end{figure}

For each subject, node and condition, six values of PE and the related uncertainty are estimated by means of the steps described in Secs.~\ref{sec:pe_assessment} and~\ref{sec:pe_uncertainty}. Figure~\ref{fig:fit_example} shows an example for a single subject belonging to the young group and a single node. The stability of the resting state is then evaluated on each of the two sets (EC and EO) of six PE assessments as follows.

First of all, stability is assumed to occur whenever PE is independent of time, i.e.\ fluctuations of PE between segments are statistically compatible with the uncertainty associated to the PE values. To this purpose, a constant value $\hat{h}$, corresponding to a horizontal line in the plot, is fitted to the data. The $p$-value corresponding to the resulting $\chi^2$ statistic is then evaluated by relying on a $\chi^2$ distribution with 5 degrees of freedom. Consequently, given a subject, node and condition, resting state is deemed to be ``unstable'' if the $p$-value falls below the Bonferroni-corrected significance threshold of $0.05 / 30 \simeq 0.0017$.

The values of the best-fit constant $\hat{h}$ for the EC, EO conditions are equal to ${5.0 \pm 0.1}$ and ${5.29 \pm 0.04}$, respectively. The $\chi^2$ test provides, for the two conditions, $p$-values equal to $2\cdot 10^{-10}$ and $0.087$, leading to the conclusion that resting state under the EC condition is unstable, whereas stability occurs under the EO condition.

\section{Stability analysis in terms of age, condition, and brain region}
\label{sec:stability_analysis}
\begin{figure}[h!]
	\centering
	\includegraphics[scale=1]{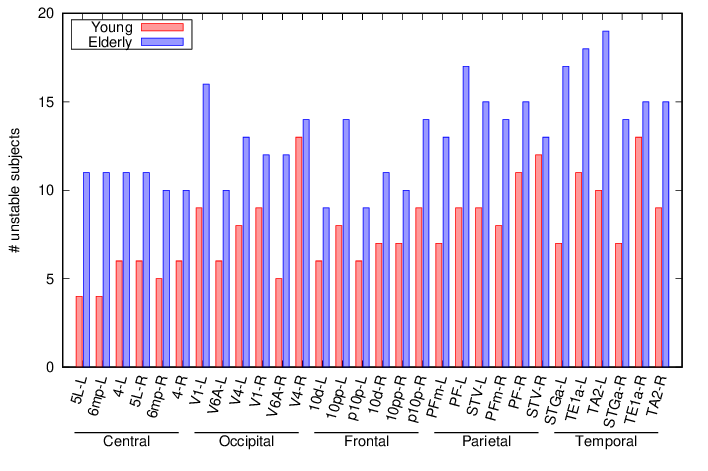}
	\caption{Number of subjects for which a node is deemed to be ``unstable'' in the EC condition.}
	\label{fig:unstable_EC}
\end{figure}

\begin{figure}[h!]
	\centering
	\includegraphics[scale=1]{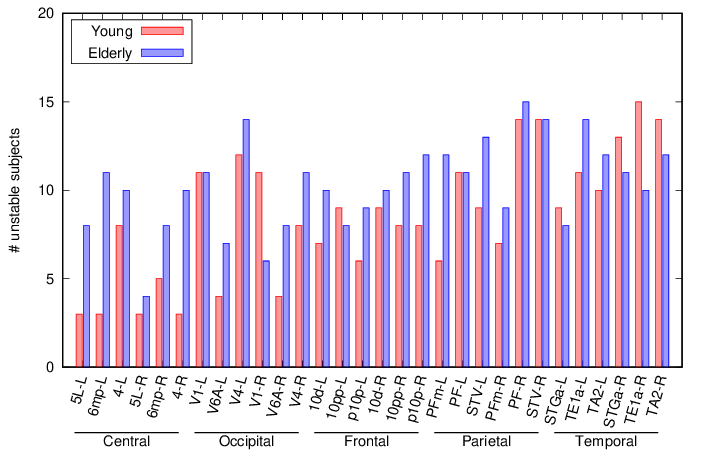}
	\caption{Number of subjects for which a node is deemed to be ``unstable'' in the EO condition.}
	\label{fig:unstable_EO}
\end{figure}

For each node, the number of subjects yielding an ``unstable'' resting state in the EC condition is shown in Fig.~\ref{fig:unstable_EC}. The analogous assessment concerning the EO condition is instead shown in Fig.~\ref{fig:unstable_EO}. In both figures, data are displayed separately for the two sets of subjects, young and elderly.

The data displayed in Figs.~\ref{fig:unstable_EC},~\ref{fig:unstable_EO} suggest that ``instability'' of resting state is not uncommon: the number of subjects in which an ``unstable'' resting state condition is detected is close to one third of the whole set of subjects. Averaging on the 30 nodes, the frequency of unstable nodes for the young group is $\simeq 26\%$ (EC) and $\simeq 28\%$ (EO), while for the elderly group it is equal to $\simeq 44\%$ (EC) and $\simeq 34\%$ (EO). In other words, as it results in Figs.~\ref{fig:unstable_EC},~\ref{fig:unstable_EO}, the number of subjects exhibiting ``instability'' is generally larger for the elderly group than for the young group, regardless of the brain location. This observation can be explained by elderly subjects tending to be more affected by fatigue.

\begin{figure}[h!]
	\centering
	\includegraphics[scale=1.2]{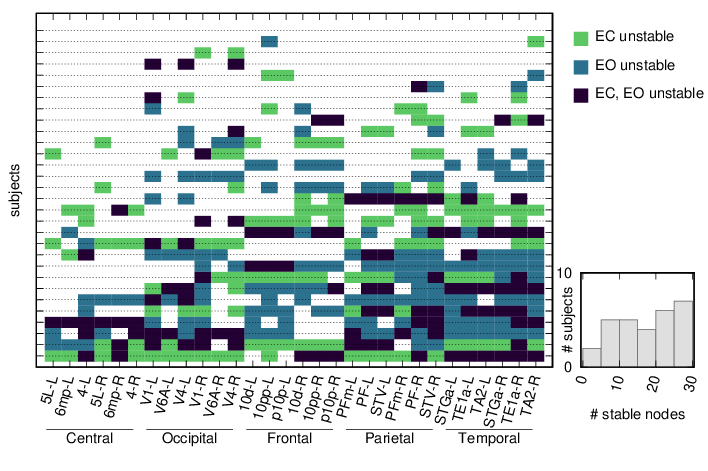}
 	\caption{Subject-node color map of resting state stability for the set of young subjects. Subjects are ordered by increasing (from bottom to top) number of stable nodes. The histogram in the inset displays the distribution of the number of stable nodes.}
 	\label{fig:subjectstat_young}
\end{figure}

\begin{figure}[h!]
	\centering
	\includegraphics[scale=1.2]{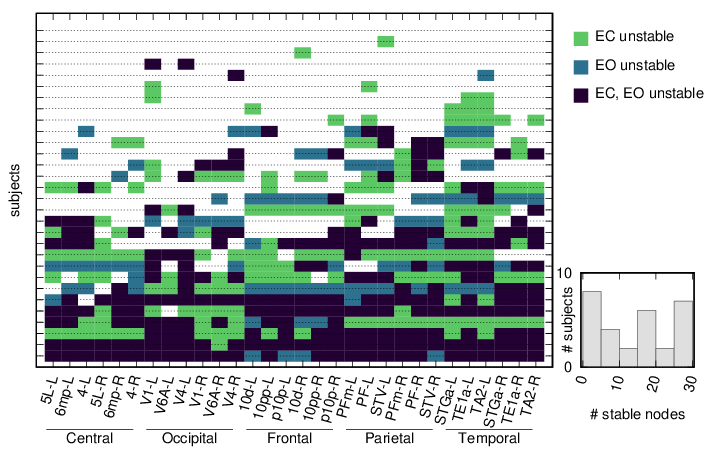}
 	\caption{Subject-node color map of resting state stability for the set of elderly subjects. Subjects are ordered by increasing (from bottom to top) number of stable nodes. The histogram in the inset displays the distribution of the number of stable nodes.}
 	\label{fig:subjectstat_elderly}
\end{figure}

One might wonder whether stability is a consistent property among nodes of the same subject, namely whether instability of resting state concerns the whole brain or, rather, few localized areas. To this purpose, Fig.~\ref{fig:subjectstat_young} shows, for the young set of subjects, a color map of stability as a function of subject and node. Specifically, to each subject-node pair, we assigned a white color if both EO and EC are deemed to be stable, and a different color in the case of unstable EC (light green), unstable EO (bluish green), or both unstable EC and EO (dark purple). The same analysis is reported in Fig.~\ref{fig:subjectstat_elderly} for the elderly set of subjects.
Besides the color map, each figure also reports an histogram of the number of subjects as a function of the number of stable nodes. While the distribution for young subjects is right-skewed, with most subjects having a number of stable nodes $\gtrsim 20$, the distribution for elderly subjects appear to be more noisy, with the majority of subjects exhibiting either a ``globally stable'' or a ``globally unstable'' resting state. This observation might reflect an underlying lower modularity in elderly subjects~\cite{NeuroImage2022}, namely the fact that brain dynamics tends to involve many areas.

\section{Complexity of stable resting states}
\label{sec:complexity_stable_resting_states}
\begin{figure}[h!]
	\centering
	\includegraphics[scale=1.2]{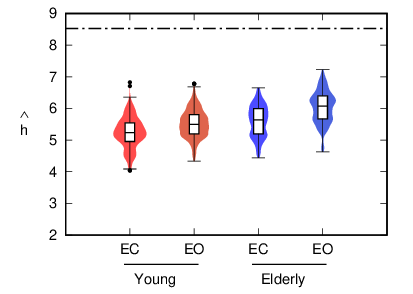}
	\caption{Distribution plots of the values of $\hat{h}$ corresponding to a ``stable'' resting state for the young-elderly and EC-EO groupings. The dash-dotted line corresponds to the maximum PE value for $m = 7$, namely $\ln 7! \simeq 8.525$. The number of $\hat{h}$ values contributing to each distribution are, from left to right: 643, 645, 507, 591.}
	\label{fig:pe_values}
\end{figure}

Whenever a combination of subject, node and condition is deemed to be stable, the corresponding $\hat{h}$ value, which is indeed an average PE, can be taken as a marker to characterize the complexity of the underlying dynamics. The distributions of $\hat{h}$ values corresponding to stable resting state are shown in Fig.~\ref{fig:pe_values}, grouped by age and condition. It is first worth highlighting that the maximum PE value with $m = 7$, corresponding to white noise time series, is $\ln 7! \simeq 8.525$: according to the data shown in Fig.~\ref{fig:pe_values}, the EEG recordings analyzed here do not correspond to a purely stochastic dynamics.
The four distributions displayed in Fig.~\ref{fig:pe_values} are comparable in width and shape, though with a slight shift towards higher PE values occurring in the elderly sets. A two-tailed $t$-test between the young-EC and elderly-EC data rejects the null-hypothesis of equal means ($p < 0.001$); the same outcome is obtained by testing the young-EO, elderly-EO sets.

\section{Discussion}
\label{sec:discussion}
In this work, we exploited EC-EO conditions as a benchmark to study resting state stability. The EC-EO paradigm is typically studied in terms of the spectral changes, most notably the presence of alpha rhythms in EC~(\cite{Barry2017}). More recently, the EC-EO paradigm was analyzed through an information-theoretic approach in two works. In the first one~\cite{QuinteroQuiroz2018}, the authors analyzed EEG recordings under the two conditions by applying a method~\cite{Masoller2015} aimed at assessing the transition probabilities of ordinal patterns (i.e.\ permutations) and determining the so-called network-averaged node entropy and asymmetry coefficient. The authors showed that the EO condition is characterized by higher entropy values, accompanied by a lower asymmetry coefficient, with respect to the EC condition. In the second paper~\cite{Vecchio2021} the authors described an approximate entropy analysis of EEG recordings recorded in the two conditions, again showing that the EO condition provides higher entropy values than the EC. Those results are in agreement with the ones obtained with the present analysis: we also found higher PE values in the EO condition. It is worth remarking that both works mentioned above carried out the respective analyses in sensor space: while this approach is computationally less demanding, one has to keep in mind that the influence of volume conduction might produce spurious results, as it was shown in the case of connectivity measures~\cite{Brunner2016,VanDeSteen2019}.

The application of information-theoretical tools to electrophysiological data is often lacking an estimation of the related uncertainty, despite the availability of asymptotic formulas~\cite{Harris1975,PRE2021b}: the significance of the quantities inferred in the analysis is typically based on averaging over a set of subjects. Conversely, the approach followed here, which relies on surrogate-based estimation of PE uncertainty, is capable of assigning an uncertainty value to each PE assessment. This capability paves the way to the possibility of drawing subject-specific quantitative conclusions from data. Besides the present application to quality control of resting state stability, such possibility is of primary importance in the development of diagnostic biomarkers based on EEG measures~\cite{Meghdadi2021} and other kinds of electrophysiological data like, for example, magnetoencephalographic recordings.

As final remarks, it is worth noting that a real time implementation of the method just depends on the hardware and software units that are used to process data. In addition, the method proposed in the present work is not limited to neuroscience: its scope can be widened to include any field where it is necessary to detect nonstationarity of a system's dynamics out of singleton time series.

\section*{Conflict of Interest Statement}
The authors have no conflicts to disclose.

\section*{Ethics approval}
EEG data used in the present work belong to the Leipzig Mind-Brain-Body ``LEMON'' database, for which the related study protocol was approved by the ethical committee at the University of Leipzig (reference number 154/13-ff). Further details are available in Ref.~\cite{Babayan2019}.

\section*{Data Availability Statement}
Raw EEG recordings used in the present work are available in the LEMON database at \url{http://fcon_1000.projects.nitrc.org/indi/retro/MPI_LEMON.html}, Ref.~\cite{LEMONweb}.

\bibliographystyle{unsrt}
\bibliography{text}

\end{document}